\documentclass[11pt]{article}
\usepackage{epsfig,latexsym}

\def\theequation{\arabic{section}.\arabic{equation}}

\renewcommand{\theequation}{\thesection.\arabic{equation}}

\global\arraycolsep=1pt
\input{tcilatex}
\begin{document}

\font\cmss=cmss10 \font\cmsss=cmss10 at 7pt \hfill \hfill IFUP-TH/02-33


\vspace{10pt}

\begin{center}
\vskip .5truecm{\Large \textbf{\vspace{10pt}}}

{\Large \textbf{``INTEGRABILITY'' OF\ RG FLOWS AND\ DUALITY\ IN THREE
DIMENSIONS IN\ THE\ 1/N EXPANSION}}

\bigskip \bigskip \vskip .5truecm

\textsl{Damiano Anselmi}

\textit{Dipartimento di Fisica E. Fermi, Universit\`{a} di Pisa, and INFN}
\end{center}

\vskip 2truecm

\begin{center}
\textbf{Abstract}
\end{center}

I study some classes of RG flows in three dimensions that are classically
conformal and have manifest $g\rightarrow 1/g$ dualities. The RG\ flow
interpolates between known (four-fermion, Wilson-Fischer, $\varphi _{3}^{6}$%
) and new interacting fixed points. These models have two remarkable
properties: $i$) the RG flow can be integrated for arbitrarily large values
of the couplings $g$ at each order of the $1/N$ expansion; $ii$) the duality
symmetries are exact at each order of the $1/N$ expansion. I integrate the
RG flow explicitly to the order $\mathcal{O}(1/N),$ write correlators at the
leading-log level and study the interpolation between the fixed points. I
examine how duality is implemented in the regularized theory and verified in
the results of this paper.

\vskip 2truecm

\vfill\eject

\section{Introduction}

\setcounter{equation}{0}

In three-dimensional quantum field theory it is often possible to derive
non-perturbative results using a large $N$ expansion. Typically, $N$ is the
number of scalar fields or fermions. To a given order of the $1/N$ expansion
infinitely many graphs can be collected by means of simple resummations and
the resulting number of effective graphs is finite. In some cases it is not
necessary to assume that the couplings $g$ of the theory are small and
expand perturbatively in $g$, but it is possible to calculate the exact $g$%
-dependence at each order of the $1/N$ expansion. Then, the computation of a
finite number of effective graphs allows us to integrate the RG flow and
study the interpolation between the fixed points. If the theories satisfy a
duality $g\leftrightarrow 1/g$, then this symmetry is implemented exactly at
each order of the $1/N$ expansion. The purpose of this paper, which is an
elaboration of the ideas of ref. \cite{largeN}, is to study a variety of
models with these features.

Specifically, I study two classes of flows. The first class of models
involves $K+1$ sets of $N_{I}$, $I=0,1,\ldots K$, fermions interacting by
means of a $\sigma $-field, that is to say a dynamical source for a
composite operator. The source $\sigma $ is called dynamical because the
functional integral is performed also over $\sigma $. The large $N$ limit is
obtained by assuming that the numbers $N_{I}$ of fermions tend to infinity
with equal velocity. At the leading order of the $1/N$ expansion the RG flow
is trivial, in the sense that the UV and IR fixed points are
indistinguishable. A nontrivial RG flow appears at the level of $\mathcal{O}%
(1/N)$ corrections. In the simplest version of these models, the fixed
points of these flows coincide with the UV fixed points of the four-fermion
models \cite{parisi,rosen} plus decoupled massless fermions. In a variant of
these models there are also fixed points of new type.

The second class of models is a scalar-field counterpart of the first class.
Sets of scalar fields interact by means of a $\sigma $-field. Due to the
radiative corrections, $\varphi ^{6}$-potential terms are generated. The RG
structure of these flows is more involved and the fixed points include known
conformal field theories (Wilson-Fischer, $\varphi _{3}^{6}$ fixed points)
and new ones.

The theories satisfy certain duality symmetries, associated with the
exchange of fermions (or scalars) belonging to different sets. The duality
symmetries are exact at each order of the $1/N$ expansion. Nevertheless, due
to some subtleties of the regularization technique \cite{largeN}, the
implementation of the duality symmetry at the regularized level needs to be
studied carefully.

The integration of the RG flow allows us to write all correlation functions
exactly in $g$ to a given order in the $1/N$ expansion. I stress again that
the expansion is not around one fixed point. Normally, an infinite
resummation is necessary to reach an interacting fixed point from a Gaussian
fixed point. Here, instead, the entire RG flow is expanded at the same time
in $1/N$ and no expansion is made in the running couplings $g$. At each
order of the $1/N$ expansion the fixed points and the flow are determined
with equal and consistent precision. Typically, both the UV\ and IR fixed
points are interacting.

\bigskip

The large $N$ expansion is crucial to make the integration of the RG flow
possible as described. To the order $\mathcal{O}(1/N)$, the integration of
the RG flow is equivalent to the resummation of the leading logs, but in
general (in QCD and non-Abelian Yang-Mills theory with matter, for example)
the resummation of the leading logs is not sufficient to connect the UV and
IR fixed points, or the high- and low-energy limits of the theory, although
it gives a sort of result beyond the perturbative expansion.

A known theory exhibiting some of the features of the models studied here is
the large $N$ $\varphi _{3}^{6}$ theory. As a classically conformal
(tricritical) theory, it is IR free and its $\mathcal{O}(1/N)$-order beta
function has a nontrivial ultraviolet fixed point \cite{pisarskietal}. The $%
\varphi _{3}^{6}$ theory, however, does not have a duality symmetry. Viewed
as a particular case of the models analysed in this paper, it belongs to a
more general family of theories which do satisfy duality.

In the study of critical phenomena, stability puts some limits on the values
of the $\varphi _{3}^{6}$ coupling \cite{moshe} and the UV\ fixed point of 
\cite{pisarskietal}, which lies in the instability region, is superseded by
a different fixed point. No attempt is made here to apply the approach of 
\cite{moshe} to the models of this paper. This study is left to a separate
research, in view of the possible interest of these theories for condensed
matter physics. In this paper I study only the tricritical domain.

Other types of ``exact'' results about RG flows exist in the literature. For
example, in four-dimensional supersymmetric theories it is often possible to
write an exact beta function \cite{NSVZ}. The NSVZ beta function, however,
depends on the subtraction scheme and the present knowledge of this scheme
does not allow us to integrate the RG flow. The NSVZ formula can be used to
obtain exact results about the fixed points \cite{noi,noi2}.

\bigskip

Since the models treated here are classically conformal, it is possible to
work with a classically conformal subtraction scheme. In practice this means
that, by default, the linear and quadratic divergences are subtracted away,
i.e. the associated renormalized dimensionful coupling constants are set to
zero. This is automatic in the framework of the dimensional regularization.
I recall that in these theories the naive dimensional regularization has to
be modified adding an evanescent, RG invariant, non-local term to the
lagrangian \cite{largeN}.

The RG flow is scheme independent to the order $O(1/N)$. Scheme dependences
appear at the order $\mathcal{O}(1/N^{2})$, but scheme-independent
correlation functions can be written to arbitrary order in the $1/N$
expansion.

The models of this paper have been inspired by the study of the
irreversibility of the RG flow \cite{proc}. This research singles out, in
even dimensions, the specialness of classically conformal theories. In odd
dimensions the investigation of the irreversibility of the RG flow is more
difficult, since we miss the tools provided, in even dimensions, by the
embedding in external gravity. It is useful to construct a large class of
classically conformal theories as a laboratory to test some ideas on this
issue. The interpretation of \cite{athm} suggests a more physical choice for
the expansion parameter, which is the scheme-invariant area of graph of the
beta function between the fixed points, defined as 
\[
\Delta a^{\prime }=\int \mathrm{d}^{3}x\,|x|^{3}\,\langle \Theta (x)\,\Theta
(0)\rangle , 
\]
where $\Theta $ is the trace of the stress tensor. The study of the quantity 
$\Delta a^{\prime }$ is left to a separate work \cite{ineq}, where the
implications on irreversibility are discussed.

\medskip

In section 2 the fermionic models are studied in detail. In section 3
correlation functions are computed explicitly and the RG\ interpolation
between the fixed points is studied. In section 4 I solve a particularly
simple, but non-trivial, limit, in which the ratios $N_{I}/N_{0}$ are sent
to zero after the large $N$ expansion. In section 5 I study the models with
two-component spinors, which have peculiar features and fixed points of new
type. Section 6 is devoted to the study of the scalar models, their beta
functions, anomalous dimensions and flows. Section 7 contains the
conclusions. In the appendix I discuss some subtleties concerning the
implementation of the duality symmetry in the framework of the modified
dimensional-regularization technique of ref. \cite{largeN}.

\section{Fermion models}

\setcounter{equation}{0}

The model we study was defined in ref. \cite{largeN} and describes $K+1$
sets of $(N_{0},\ldots ,N_{K})$ fermions interacting by means of a $\sigma $%
-field. The lagrangian is 
\begin{equation}
\mathcal{L}=\sum_{I=0}^{K}\sum_{i=1}^{N_{I}}\overline{\psi }%
_{i}^{I}(\partial \!\!\!\slash +\lambda _{I}\sigma )\psi _{i}^{I}.
\label{lagrangian}
\end{equation}
To simplify various arguments, I first choose four-component complex
spinors. The generalization to models with two-component spinors, which have
fixed points of new type, is discussed later. I work in the Euclidean
framework, where $\psi $ and $\psi ^{\dagger }$ are independent from each
other. The gamma matrices are Hermitian and $\overline{\psi }=\psi ^{\dagger
}\gamma _{3}$. A basis with $\gamma _{1,3}^{T}=\gamma _{1,3}$ and $\gamma
_{2}^{T}=-\gamma _{2}$ is used for charge conjugation (see below). The
matrices $\gamma _{4}$ and $\gamma _{5}$ of the four-dimensional Dirac
algebra act on our spinors as flavour matrices. Since the fundamental spinor
in three dimensions is real (Majorana) and has two components, a set of $%
N_{f}$ massless four-component complex spinors has a flavour symmetry $%
O(4N_{f})$, while a set of $N_{f}$ massive four-component complex spinors
has symmetry $O(2N_{f})\otimes O(2N_{f}).$ For other details see for example 
\cite{rosen}. The field $\sigma $ is a sort of dynamical fermion mass, but
our fermions are otherwise massless. The $\sigma $ expectation value is zero
in the classically conformal phase.

\bigskip

\textbf{Symmetries.} The theory is manifestly invariant under reflection
positivity (Hermitian conjugation combined with a sign-change of the
``time'' coordinate $x_{3}$), parity, charge conjugation and time-reversal.
The P, C and T transformations are 
\begin{eqnarray*}
P_{1} &:&\qquad x_{1}\rightarrow -x_{1},\qquad x_{2,3}\rightarrow
x_{2,3},\qquad \psi _{i}^{I}\rightarrow \gamma _{1}\gamma _{4}\psi
_{i}^{I},\qquad \sigma \rightarrow \sigma \ ; \\
C &:&\qquad x_{1,2,3}\rightarrow x_{1,2,3},\qquad \psi _{i}^{I}\rightarrow
\gamma _{2}\left( \overline{\psi }_{i}^{I}\right) ^{T},\qquad \sigma
\rightarrow \sigma \ ; \\
T &:&\qquad x_{3}\rightarrow -x_{3},\qquad x_{1,2}\rightarrow x_{1,2},\qquad
\psi _{i}^{I}\rightarrow \gamma _{3}\gamma _{4}\psi _{i}^{I},\qquad 
\overline{\psi }_{i}^{I}\rightarrow \overline{\psi }_{i}^{I}\gamma
_{4}\gamma _{3},\qquad \sigma \rightarrow \sigma .
\end{eqnarray*}
Further symmetris are obtained replacing $\gamma _{4}$ with $\gamma _{5}$.
It is understood that, unless otherwise stated, the transformation of $%
\overline{\psi }$ is obtained from the one of $\psi $ as if $\psi ^{\dagger
} $ were the adjoint of $\psi $. In the list above, only time reversal
requires to transform $\psi $ and $\psi ^{\dagger }$ independently.

The theory is also invariant under the chiral transformation 
\begin{equation}
\psi _{i}^{I}\rightarrow \gamma _{4}\psi _{i}^{I},\qquad \quad \sigma
\rightarrow -\sigma ,  \label{ch}
\end{equation}
and the analogous transformation with $\gamma _{4}\rightarrow \gamma _{5}.$
A consequence of chiral invariance is that the graphs with an odd number of
external $\sigma $-legs vanish. In particular, no vertex $\sigma ^{3}$ is
generated by renormalization.

The global symmetry group of the theory (\ref{lagrangian}) is $%
\bigotimes_{I=0}^{K}\left( O(2N_{I})\otimes O(2N_{I})\right) $, enhanced at
the fixed points as discussed below.

Further simmetries are the $g\rightarrow 1/g$ dualities, which I present
after fixing some other notation.

If we perform a transformation 
\begin{equation}
\psi _{i}^{\bar{I}}\rightarrow \gamma _{4}\psi _{i}^{\bar{I}},\qquad \sigma
\rightarrow -\sigma ,  \label{chsign}
\end{equation}
for some $\bar{I}$ only, we change the sign of the corresponding coupling $%
\lambda _{\bar{I}}$. So, we can assume that all of the $\lambda _{I}$s are
non-negative.

\bigskip

\textbf{Renormalization.} The field $\sigma$ has no propagator at the
classical level, but has a propagator proportional to $1/\sqrt{k^2}$ at the
quantum level, generated by the fermion bubble. The theory is defined using
the Parisi large $N$ approach of ref. \cite{parisi}. The field $\sigma$ has,
by definition, the same dimension as the fermions, namely $(d-1)/2$.

Since the theory is classically conformal, we can choose the classically
conformal subtraction scheme, in which the linear divergences are just
ignored, which is equivalent to say that the renormalized fermion mass and
the renormalized expectation value of $\sigma$ vanish.

The dimensional regularization ($d=3-\varepsilon $) and the minimal
subtraction scheme are used. Having chosen doublets of complex spinors, we
can set the trace of an arbitrary odd number of gamma matrices to zero. The
symmetries listed in the previous paragraph are manifestly preserved at the
quantum level.

It was shown in \cite{largeN} that the naive dimensional-regularization
technique does not work properly, because the fermion bubble generates a $%
\sigma $-propagator proportional to $1/(k^{2})^{2-d/2}$, which is
responsible for the appearance of $\Gamma [0]$s at the subleading orders. To
properly regularize the theory, it is necessary to manually convert the $%
\sigma $-propagator to $1/\sqrt{k^{2}}$. This can be achieved adding an
RG-invariant evanescent non-local term to the lagrangian.

We choose a preferred set of fermions, say $I=0$, as a reference for the
large $N$ expansion. We define $\lambda \equiv \lambda _{0}$, $N\equiv N_{0}$%
, $\psi _{i}\equiv \psi _{i}^{0}$, etc. We take $N$ large, with $%
r_{J}=N_{J}/N_{0}$, $\alpha _{J}=\lambda _{J}^{2}/\lambda _{0}^{2}$ and 
\begin{equation}
\lambda ^{2}N=4  \label{fixed}
\end{equation}
fixed. The numerical value in (\ref{fixed}) is conventional. (\ref{fixed})
holds only in $d=3$, because for $d=3-\varepsilon $ the dimensionality of $%
\sigma $ and the identity $\lambda _{\mathrm{B}}=\lambda \mu ^{\varepsilon
/2}$ give an evanescent $\lambda $ beta function, equal to 
\[
\beta _{\lambda }=-{\frac{1}{2}}\varepsilon . 
\]

The couplings $\alpha _{J}$ are exactly dimensionless. We define their
renormalization constants by $(\alpha _{I})_{\mathrm{B}}=\alpha _{I}Z_{I}$.
The renormalized lagrangian reads 
\[
\mathcal{L}=\sum_{i=1}^{N}Z_{\psi }\overline{\psi }_{i}(\partial \!\!\!%
\slash +\lambda \mu ^{\varepsilon /2}Z_{\sigma }^{1/2}\sigma )\psi
_{i}+\sum_{I=1}^{K}\sum_{i=1}^{N_{I}}Z_{\psi }^{I}\overline{\psi }%
_{i}^{I}(\partial \!\!\!\slash +\lambda _{I}\mu ^{\varepsilon
/2}Z_{I}^{1/2}Z_{\sigma }^{1/2}\sigma )\psi _{i}^{I}. 
\]

\bigskip

\textbf{Dualities.} The duality symmetries of the theory are related to the
exchanges of fermions belonging to different sets \cite{largeN}. First of
all, the theory is evidently symmetric under the exchange of $\bar{I}$ with $%
\bar{J}$ with $\bar{I},\bar{J}>0$, which means $\psi _{\bar{I}%
}\leftrightarrow \psi _{\bar{J}}$, $\alpha _{\bar{I}}\leftrightarrow \alpha
_{\bar{J}}$ and $r_{\bar{I}}\leftrightarrow r_{\bar{J}}$.

Let us now exchange the $\bar I$th set of fermions, $\bar{I}>0$, with the
zeroth set of fermions. The theory is symmetric under the duality
transformation 
\begin{eqnarray}
\psi_{\bar I}\leftrightarrow \psi_0,\qquad \psi_J\rightarrow\psi_J,\qquad
N_{\bar I}\leftrightarrow N_0,\qquad N_J\rightarrow N_J,\qquad
\lambda_0\rightarrow {\frac{\lambda_0}{\sqrt{r_{\bar I}}}},  \nonumber \\
\sigma\rightarrow\sigma{\frac{\lambda_{\bar I}}{\lambda_0}}\sqrt{r_{\bar I}}%
, \qquad\alpha_{\bar I}\rightarrow {\frac{1}{\alpha_{\bar I}}},\qquad
\alpha_J\rightarrow {\frac{\alpha_J}{\alpha_{\bar I}}},\qquad r_{\bar
I}\rightarrow {\frac{1}{r_{\bar I}}},\qquad r_J\rightarrow {\frac{r_J}{%
r_{\bar I}}},  \label{dual}
\end{eqnarray}
with $J\neq \bar I,0$. The normalization condition (\ref{fixed}) is duality
invariant. The transformation (\ref{dual}) will be called ``the $\bar I$th
duality transformation''.

I now prove that, without loss of generality, we can take all of the $\alpha 
$s to have values between zero and one. Suppose that $\alpha _{1}>1$. (If $%
\alpha _{1}\leq 1$ then skip to $\alpha _{2}$.) The first duality
transformation changes $\alpha _{1}$ to $\alpha _{1}^{\prime }=1/\alpha
_{1}<1$ and $\alpha _{I}$ to $\alpha _{I}^{\prime }=\alpha _{I}/\alpha _{1}$
for $I>1$. Now suppose that $\alpha _{2}^{\prime }>1$. (If $\alpha
_{2}^{\prime }\leq 1$ then skip to $\alpha _{3}^{\prime }$.) The second
duality transformation changes $\alpha _{1}^{\prime }<1$ to $\alpha
_{1}^{\prime \prime }=\alpha _{1}^{\prime }/\alpha _{2}^{\prime }<1$, $%
\alpha _{2}^{\prime }$ to $\alpha _{2}^{\prime \prime }=1/\alpha
_{2}^{\prime }<1$ and $\alpha _{I}^{\prime }$ to $\alpha _{I}^{\prime \prime
}=\alpha _{I}^{\prime }/\alpha _{2}^{\prime }$, for $I>2$. The important
point is that the duality transformation sends couplings with values $\leq 1$
into couplings with values $\leq 1$. Clearly, after at most $K$ steps we
have $0\leq \alpha _{I}\leq 1$ for every $I$.

In conclusion, it is not restrictive to assume 
\[
0\leq \alpha_I\leq 1,\qquad \forall I, 
\]
which will be understood henceforth.

\bigskip

\textbf{Fixed points.} The conformal fixed points of the RG flow are the
models where $(\alpha _{1},\ldots \alpha _{K})$ is any string made of $0$s
and $1$s. These theories are always the direct product 
\begin{equation}
\Psi _{N_{f}}\otimes F\Sigma _{N_{\sigma }},  \label{fixedF}
\end{equation}
of a theory $\Psi _{N_{f}}$ of $N_{f}$ free massless fermions and the
interacting conformal field theory (called $\sigma _{N_{\sigma }}$ in ref. 
\cite{largeN}) identified by the lagrangian 
\begin{equation}
\mathcal{L}=\sum_{i=1}^{N_{\sigma }}\overline{\psi }_{i}(\partial \!\!\!%
\slash +\lambda \sigma )\psi _{i}.  \label{unit}
\end{equation}
For practical convenience, this theory will be called $F\Sigma _{N_{\sigma
}} $ henceforth. Here $N_{f}$ is the total number of decoupled fermions ($%
\alpha _{I}=0$) and $N_{\sigma }$ is the total number of fermions coupled to 
$\sigma $ with $\alpha _{I}=1$. The theory (\ref{fixedF}) has an enhanced
flavor symmetry group $O(4N_{f})\otimes O(2N_{\sigma })\otimes O(2N_{\sigma
})$. The conformal field theory $F\Sigma _{N_{\sigma }}$ coincides with the
UV\ fixed point of the four-fermion models 
\[
\sum_{i=1}^{N}\overline{\psi }_{i}\partial \!\!\!\slash\psi _{i}-{\frac{1}{M}%
}\left( \sum_{i=1}^{N}\overline{\psi }_{i}\psi _{i}\right) ^{2}. 
\]
Following \cite{largeN} these conformal field theories can be formulated 
\textsl{per se}, in a classically conformal framework, with no reference to
a flow originating them as its large- or small-distance limit.

\bigskip

\textbf{The zeroth order of the large $N$ expansion.} The $\sigma $%
-propagator is dynamically generated by the one-loop fermion bubble (graph ($%
a$) of Fig. 1) and reads 
\begin{equation}
\langle \sigma (k)\,\sigma (-k)\rangle ={\frac{1}{\sqrt{k^{2}}}}D(\alpha
_{J},r_{J}),\qquad D^{-1}(\alpha _{J},r_{J})=1+\sum_{J=1}^{K}\alpha
_{J}r_{J}.  \label{ss}
\end{equation}

\begin{figure}[tbp]
\centerline{\epsfig{figure=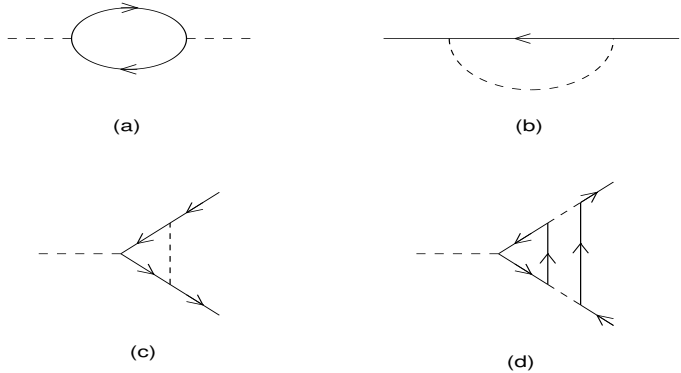,height=5cm,width=9cm}}
\caption{Order $\mathcal{O}(1/N)$ graphs}
\end{figure}

To the leading order of the large $N$ expansion the quantum action reads 
\begin{equation}
\Gamma =\int \mathrm{d}^{d}x\,\sum_{I=0}^{K}\sum_{i=1}^{N_{I}}\overline{\psi 
}_{i}^{I}\partial \!\!\!\slash\psi _{i}^{I}+\int {\frac{\mathrm{d}^{d}k}{%
(2\pi )^{d}}}|\sigma (k)|^{2}\sqrt{k^{2}}D^{-1}(\alpha _{J},r_{J}),
\label{effa}
\end{equation}
This is a Gaussian theory, with a non-local kinetic term for $\sigma $. The
couplings are inert and no flow takes place. In this limit, the UV and IR
fixed points are indistinguishable, and the constant $D(\alpha _{J},r_{J})$
can be reabsorbed in $\sigma $. It is immediate to check that the
zeroth-order expression (\ref{effa}) of $\Gamma $ is invariant under the
duality transformation (\ref{dual}).

\bigskip

\textbf{The first order of the large $N$ expansion.} The RG flow can be
inspected computing the $\mathcal{O}(1/N)$ corrections. The relevant graphs
are the one-loop fermion self-energy and the one-loop vertex (see Fig. 1) 
\cite{largeN}. A two-loop graph ($d$) containing a fermion triangle could in
principle contribute to the vertex, but it is zero because of the
conservation of chirality.

After resumming the fermion bubbles, the number of relevant graphs becomes
finite at each order of the $1/N$ expansion. This means that the $\alpha
_{I} $-dependence can be determined exactly, order by order in the $1/N$
expansion. Moreover, this knowledge is sufficient to connect the fixed
points, so that it is possible to write $\alpha _{I}$-exact beta functions
and correlation functions to any given order of the $1/N$ expansion. The
order-$\mathcal{O}(1/N)$ expressions of the beta functions and anomalous
dimensions allow us to resum the leading logs (see next section).

The order-$\mathcal{O}(1/N)$ renormalization constants can be written using
the results of \cite{largeN} and read 
\begin{eqnarray}
Z_{\psi }^{I} &=&1-{\frac{2}{3\pi ^{2}N\varepsilon }}\alpha _{I}\mu
^{-\varepsilon }D(\alpha _{J},r_{J}),\qquad Z_{\sigma }=1+{\frac{16}{3\pi
^{2}N\varepsilon }}\mu ^{-\varepsilon }D(\alpha _{J},r_{J}),  \label{zetas}
\\
Z_{I} &=&1+{\frac{16}{3\pi ^{2}N\varepsilon }}(\alpha _{I}-1)\mu
^{-\varepsilon }D(\alpha _{J},r_{J}).  \nonumber
\end{eqnarray}
These expressions are correct also for $I=0$. The factors $D(\alpha
_{J},r_{J})$ are brought by the $\sigma $-propagators. The factor $\mu
^{-\varepsilon }$ should be understood as a short form for $\lambda _{%
\mathrm{B}}^{2}N\mu ^{-\varepsilon }/4$, so that, using (\ref{fixed}), 
\[
\lim_{\varepsilon \rightarrow 0}\frac{\mathrm{d}}{\mathrm{d}\ln \mu }\left[ 
\frac{1}{\varepsilon }\left( \frac{\lambda _{\mathrm{B}}^{2}N\mu
^{-\varepsilon }}{4}\right) ^{m}\right] =-m. 
\]
We immediately derive anomalous dimensions and beta functions: 
\begin{eqnarray}
\gamma _{\psi }^{I} &=&{\frac{1}{2}}{\frac{\mathrm{d}\ln Z_{\psi }^{I}}{%
\mathrm{d}\ln \mu }}={\frac{1}{3\pi ^{2}N}}\alpha _{I}D(\alpha
_{J},r_{J}),\qquad \gamma _{\sigma }={\frac{1}{2}}{\frac{\mathrm{d}\ln
Z_{\sigma }}{\mathrm{d}\ln \mu }}=-{\frac{8}{3\pi ^{2}N}}D(\alpha
_{J},r_{J}),  \nonumber \\
\beta _{I}(\alpha ) &=&-\alpha _{I}{\frac{\mathrm{d}\ln Z_{I}}{\mathrm{d}\ln
\mu }}={\frac{16}{3\pi ^{2}N}}{\alpha _{I}(\alpha _{I}-1)}D(\alpha
_{J},r_{J}).  \label{betas}
\end{eqnarray}
The duality (\ref{dual}) can be checked explicitly in these formulas. The
only non-straightforward step in this check is the trasformation of $\gamma
_{\sigma }$: 
\begin{equation}
\gamma _{\sigma }\rightarrow \gamma _{\sigma }-{\frac{1}{2\alpha _{\bar{I}}}}%
\beta _{\bar{I}}\equiv \widetilde{\gamma }_{\sigma }^{\bar{I}},  \label{gs}
\end{equation}
which is implied by $\sigma \rightarrow \sigma \sqrt{r_{\bar{I}}}\lambda _{%
\bar{I}}/\lambda _{0}$. We are going to check duality also in the exact
expressions of the correlation functions of the next section. From the
duality relation (\ref{gs}) we recover the formula for the beta function
derived in \cite{largeN}, 
\[
\beta _{\bar{I}}=2\alpha _{\bar{I}}\left( \gamma _{\sigma }-\widetilde{%
\gamma }_{\sigma }^{\bar{I}}\right) , 
\]
which is exact.

\bigskip

\textbf{Solution of the RG equations.} The RG flow can be integrated
immediately in the space of couplings. We have 
\begin{equation}
{\frac{\mathrm{d}\alpha _{I}}{\alpha _{I}(\alpha _{I}-1)}}={\frac{\mathrm{d}%
\alpha _{J}}{\alpha _{J}(\alpha _{J}-1)}}  \label{flow}
\end{equation}
for every $I$ and $J$. The solution of (\ref{flow}) is 
\begin{equation}
{\frac{1}{\alpha _{I}}}-1=C_{IJ}\left( {\frac{1}{\alpha _{J}}}-1\right) ,
\label{solu}
\end{equation}
with no summation over $J$. The matrix $C_{IJ}$ is a matrix of arbitrary
constants. Restricting to $0\leq \alpha _{I}\leq 1$, self-consistency of (%
\ref{solu}) demands that $C_{IJ}$ satisfies 
\[
C_{IJ}\geq 0,\qquad C_{IJ}C_{JL}=C_{IL},\qquad C_{IJ}={\frac{1}{C_{JI}}}, 
\]
where, again, no summation on $J$ is understood. The most general solution
to these constraints is determined by the constants $C_{I}\equiv C_{1I}$
with $C_{IJ}=C_{J}/C_{I}$. We can write 
\begin{equation}
\alpha _{I}={\frac{C_{I}\alpha }{1+(C_{I}-1)\alpha }},  \label{ci}
\end{equation}
where $\alpha \equiv \alpha _{1}$. The choices $C_{I}=0,1,\infty $ for some $%
I$s correspond to freeze some sets of fermions or glue them to other sets of
fermions.

The dependence on the energy is obtained solving the RG equation 
\[
\beta (\alpha )={\frac{16}{3\pi ^{2}N}}{\frac{\alpha (\alpha -1)}{%
1+r_{1}\alpha +\sum_{J=2}^{K}C_{J}r_{J}\alpha /\left( 1+(C_{J}-1)\alpha
\right) }}=-{\frac{\mathrm{d}\alpha }{\mathrm{d}t}}, 
\]
where $t=\ln |x|\mu $ and $|x|$ is the overall scale of the correlation
function. This equation can always be solved implicitly in the form $%
t=t(\alpha )$. In general, the function $t(\alpha )$ can be inverted only
numerically, but in the simplest case ($K=1$, $r_{1}=1$) it can be inverted
also explicitly. The solution is then 
\[
\alpha (1/|x|)=1+u-\sqrt{u(u+2)},\qquad u={\frac{(1-\alpha (\mu ))^{2}}{%
2\alpha (\mu )}}\left( |x|\mu \right) ^{-16/(3\pi ^{2}N)}. 
\]
We have $\alpha (1/|x|)\rightarrow 0$ for $|x|\rightarrow 0$ (ultraviolet)
and $\alpha (1/|x|)\rightarrow 1$ for $|x|\rightarrow \infty $ (infrared).
For $K=1$ and $r_{1}=r$ generic we have 
\[
\left( |x|\mu \right) ^{-16/(3\pi ^{2}N)}={\frac{\alpha (\mu )}{\alpha
(1/|x|)}}\left( {\frac{1-\alpha (1/|x|)}{1-\alpha (\mu )}}\right) ^{r+1}. 
\]

\section{Correlation functions in fermion models}

\setcounter{equation}{0}

In this section I show how the integration of the RG flow allows us to write
all correlation functions exactly in $\alpha $ to any given order in $1/N$.
Here it is understood that $\ln \mu \sim N$, because the powers of $(\ln \mu
)/N$ are resummed. In particular, to write the correlation functions at the
leading-log level, it is sufficient to know the beta function and
renormalization constants to the order $1/N$. For concreteness, I choose $%
K=1 $ and $r=1$. This model has two sets of fermions, $\psi ^{0}$ and $\psi
^{1}$, and one coupling constant $\alpha $.

\bigskip

\textbf{The $\sigma $ two-point function.} Using the Callan-Symanzik
equations we know that the $\sigma $ two-point function has the form 
\[
\langle \sigma (x)\,\sigma (0)\rangle =Z_{\sigma }^{-1}(\alpha
(1/|x|),\alpha (\mu )){\frac{A(\alpha (1/|x|))}{2\pi ^{2}|x|^{2}}}. 
\]
At the leading-log level, we need $\gamma _{\sigma }$ to $\mathcal{O}(1/N)$,
given by (\ref{betas}), and $A$ to $\mathcal{O}(1)$, which is just $%
1/(1+\alpha )$, from (\ref{ss}). We find 
\[
Z_{\sigma }(\alpha (1/|x|),\alpha (\mu ))=\exp \left( -2\int_{\ln \mu
}^{-\ln |x|}\gamma _{\sigma }(\mu ^{\prime })\ \mathrm{d}\ln \mu ^{\prime
}\right) ={\frac{1-\alpha (1/|x|)}{\alpha (1/|x|)}}{\frac{\alpha (\mu )}{%
1-\alpha (\mu )}} 
\]
and therefore 
\begin{equation}
\langle \sigma (x)\,\sigma (0)\rangle ={\frac{1}{2\pi ^{2}|x|^{2}}}{\frac{%
\alpha (1/|x|)}{1-\alpha ^{2}(1/|x|)}}{\frac{1-\alpha (\mu )}{\alpha (\mu )}}%
={\frac{1}{4\pi ^{2}|x|^{2}}}{\frac{1}{\sqrt{u(u+2)}}}{\frac{1-\alpha (\mu )%
}{\alpha (\mu )}}.  \label{exactf}
\end{equation}
It is amusing to check the invariance of this correlation function under the
duality transformation (\ref{dual}). In doing this, remember that two $%
\sigma $s get a factor $\alpha (\mu )$ in the transformation.

The exact formula (\ref{exactf}) exhibits the dependence on the reference
scale $\mu $, which survives at both critical points and is different in the
UV and IR limits. The anomalous dimension varies from $1-8/(3\pi ^{2}N)$ in
the ultraviolet, to $1-4/(3\pi ^{2}N)$ in the infrared, as expected. The UV
and IR behaviors are 
\[
\langle \sigma (x)\,\sigma (0)\rangle \sim {\frac{1}{2\pi
^{2}|x|^{2(1-8/(3\pi ^{2}N))}}}{\frac{\mu ^{16/(3\pi ^{2}N)}}{1-\alpha (\mu )%
}}\text{\qquad and \qquad }\sim {\frac{1}{4\pi ^{2}|x|^{2(1-4/(3\pi ^{2}N))}}%
}{\frac{\mu ^{8/(3\pi ^{2}N)}}{\sqrt{\alpha (\mu )}}}, 
\]
respectively.

\bigskip

\textbf{The fermion two-point functions.} With the same procedure, the
fermion two-point functions are found to be 
\begin{eqnarray*}
\langle \overline{\psi }^{0}(x)\,\psi ^{0}(0)\rangle &=&{\frac{x\!\!\!\slash%
}{4\pi |x|^{3}}}\left[ {\frac{1-\alpha (1/|x|)}{\alpha (1/|x|)}}{\frac{%
\alpha (\mu )}{1-\alpha (\mu )}}\right] ^{1/8}, \\
\langle \overline{\psi }^{1}(x)\,\psi ^{1}(0)\rangle &=&{\frac{x\!\!\!\slash%
}{4\pi |x|^{3}}}\left[ {\frac{1-\alpha (1/|x|)}{1-\alpha (\mu )}}\right]
^{1/8}.
\end{eqnarray*}
There formulas are interchanged by the duality transformation (\ref{dual})
(observe that under duality $\alpha (1/|x|)\rightarrow 1/\alpha (1/|x|)$ and 
$\alpha (\mu )\rightarrow 1/\alpha (\mu )$).

\bigskip

\textbf{The $\sigma ^{2}$ two-point function.} As a further example, we
study the two-point function of the composite operator $\sigma ^{2}$. This
operator has dimension 2, but to the order $1/N$ it does not mix with the
fermion bilinears $\overline{\psi }^{0}\psi ^{0}$ and $\overline{\psi }%
^{1}\psi ^{1}$ (it is sufficient to consider only one such bilinear, since
the $\sigma $ field equation relates the two to each other). The calculation
of the order-$\mathcal{O}(1/N)$ anomalous dimension of $\sigma ^{2}$
requires the evaluation of the divergent parts of two two-loop diagrams with
a fermion loop (see Fig. 2). Since both the fermions with $I=0,1$ circulate
in the loops, the result is multiplied by $1+\alpha ^{2}$. Since there are
two $\sigma $-propagators in these diagrams, the result has also a factor $%
1/(1+\alpha )^{2}$ . We obtain, taking into account of the combinatorial
factor and the number of ways in which the fermions can circulate in the
loop, 
\[
Z_{\sigma ^{2}}^{-1}Z_{\sigma }=1+{\frac{4}{N\pi ^{2}\varepsilon }}\mu
^{-2\varepsilon }{\frac{1+\alpha ^{2}}{(1+\alpha )^{2}}} 
\]
and therefore 
\[
\gamma _{\sigma ^{2}}={\frac{8}{3N\pi ^{2}}}{\frac{1-2\alpha +3\alpha ^{2}}{%
(1+\alpha )^{2}}}. 
\]

\begin{figure}[tbp]
\centerline{\epsfig{figure=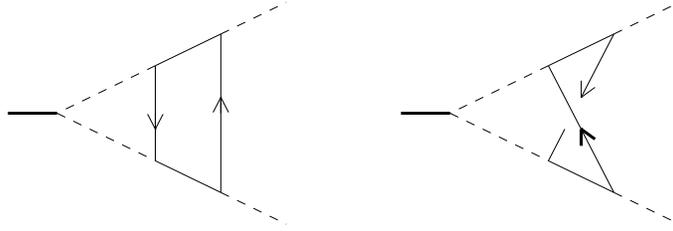,height=3cm,width=9cm}}
\caption{Renormalization of the operator $\sigma^2$}
\end{figure}

Integrating the RG equations we find 
\[
\langle \sigma ^{2}(x)\,\sigma ^{2}(0)\rangle =Z_{\sigma ^{2}}^{-2}(\alpha
(1/|x|),\alpha (\mu )){\frac{A_{2}(\alpha (1/|x|))}{8\pi ^{4}|x|^{4}}}={%
\frac{\sqrt{u(u+2)}}{4\pi ^{4}|x|^{4}}}{\frac{\alpha (\mu )}{(1-\alpha
^{2}(\mu ))(1+\alpha (\mu ))^{2}}}. 
\]
The function $A_{2}(\alpha )$ can be read from the order-$\mathcal{O}(1)$
expression of the correlation function, which is the $\sigma $-bubble. We
have $A_{2}(\alpha )=1/(1+\alpha )^{2}$.

Again, the correlation function transforms correctly under the duality (\ref
{dual}).

In the ultraviolet and infrared limits we have 
\[
{\frac{1}{8\pi ^{4}|x|^{2(2+8/(3\pi ^{2}N))}}}{\frac{\mu ^{-16/(3\pi
^{2}N)}(1-\alpha (\mu ))}{(1+\alpha (\mu ))^{3}}\qquad }\text{and\qquad }{%
\frac{1}{4\pi ^{4}|x|^{2(2+4/(3\pi ^{2}N))}}}{\frac{\mu ^{-8/(3\pi ^{2}N)}%
\sqrt{\alpha (\mu )}}{(1+\alpha (\mu ))^{3}},} 
\]
respectively.

In more generality, the anomalous dimension of the operator $\sigma ^{2}$ is 
\[
\gamma _{\sigma ^{2}}={\frac{8}{3\pi ^{2}N}}D^{2}(\alpha _{J},r_{J})\left(
1+\sum_{J=1}^{K}\alpha _{J}r_{J}(3\alpha _{J}-2)\right) 
\]
and its transformation under the $\bar{I}$th duality reads $\gamma _{\sigma
^{2}}\rightarrow \gamma _{\sigma ^{2}}-\beta _{\bar{I}}{/}\alpha _{\bar{I}}.$%
\bigskip

\textbf{Other correlation functions.} The integration of the RG flow solves
the $\alpha $-dependence and allows us to write the correlation functions
exactly in $\alpha $ to any given order in $1/N$. In this expansion, if $s$
denotes the overall scale of the correlation function, $\ln s\mu $ is
considered of order $N$ and the powers of $(\ln s\mu )/N$ are resummed. In
the previous paragraph I have considered two-point functions explicitly, but
the procedure is general.

Let us consider the three-point function 
\[
\langle \sigma ^{2}(x)\,\sigma ^{2}(y)\,\sigma ^{2}(0)\rangle \equiv
G(|x|,|y|,|x-y|). 
\]
This correlation function can be more conveniently expressed in terms of the
variables 
\[
s=|x|+|y|+|x-y|,\quad \nu _{1}={\frac{|x|}{|y|}}+{\frac{|y|}{|x|}},\quad \nu
_{2}={\frac{(x-y)^{2}}{|x||y|}}. 
\]
The variables $\nu _{1,2}$ are scale-independent and $s$ is the overall
scale. We can write 
\[
G(s,\nu _{1},\nu _{2},\alpha (\mu ),N)={\frac{1}{s^{3}}}\sum_{k}{\frac{(\ln
s\mu )^{k}}{N^{k}}}f_{k}(\nu _{1},\nu _{2},\alpha (\mu ),N). 
\]
For the leading-log approximation it is sufficient to know the functions $%
f_{k}$ to the leading order. The Callan-Symanzik equations give 
\[
G(s,\nu _{1},\nu _{2},\alpha (\mu ),N)={\frac{1}{s^{3}}}Z_{\sigma
^{2}}^{-3}(\alpha (1/s),\alpha (\mu ))G_{\mathrm{red}}(\nu _{1},\nu
_{2},\alpha (1/s),N) 
\]
and it is sufficient to compute $G_{\mathrm{red}}(\nu _{1},\nu _{2},\alpha
(1/s),N)$ to the leading order in $1/N$. We have 
\begin{equation}
\langle \sigma ^{2}(x)\,\sigma ^{2}(y)\,\sigma ^{2}(0)\rangle ={\frac{1}{\pi
^{6}|x||y||x-y|}}\left( 4u(s)(u(s)+2)\right) ^{3/4}\left( {\frac{\alpha (\mu
)}{(1-\alpha ^{2}(\mu ))(1+\alpha (\mu ))^{2}}}\right) ^{3/2},  \label{s}
\end{equation}
where $u(s)=(s\mu )^{-16/(3\pi ^{2}N)}(1-\alpha (\mu ))^{2}/(2\alpha (\mu ))$%
. If we choose a different definition of the variable $s$, formula (\ref{s})
is unaffected at the level of leading logs.

\section{The $r=0$ limit}

\setcounter{equation}{0}

A particularly simple limit is the case in which the $r_{I}$s are zero.
Setting $r_{I}=0$, $I=1,\ldots K$, after expanding in $1/N$ is not
equivalent to kill the fermions with $I=1,\ldots K$. The large $N$ and $%
r_{I}\rightarrow 0$ limits conflict and leave a non-trivial remnant. When $%
r_{I}=0$, however, duality is not manifest. We can see this immediately from
the expression of the beta function:

\[
\beta _{I}(\alpha )={\frac{16}{3\pi ^{2}N}}{\alpha _{I}(\alpha _{I}-1)}. 
\]
The RG flow interpolates between $\alpha =0$ and $\alpha =1$, but cannot
reach the fixed point at $\alpha =\infty $. To reach $\alpha =\infty $ we
have to perform a dual limit in which some or all of the $r_{I}$s tend to
infinity. The solution reads 
\[
\alpha _{I}(1/|x|)={\frac{\alpha (\mu )}{\alpha (\mu )+(1-\alpha (\mu
))C_{I}^{-1}(|x|\mu )^{-16/(3\pi ^{2}N)}}}, 
\]
where $\alpha \equiv \alpha _{1}$ and $C_{1}=1$. Formula (\ref{ci}) still
holds. The anomalous dimensions are 
\[
\gamma _{\psi }^{I}={\frac{1}{3\pi ^{2}N}}\alpha _{I},\qquad \gamma _{\sigma
}=-{\frac{8}{3\pi ^{2}N}},\qquad \gamma _{\sigma ^{2}}={\frac{8}{3\pi ^{2}N}}%
. 
\]
$\gamma _{\sigma }$ and $\gamma _{\sigma ^{2}}$ do not depend on the energy.
The other results can be generalized straightforwardly to this case.

\section{Models with two-component spinors and new fixed points}

Let us now consider a variant of the model (\ref{lagrangian}) in which the
spinors have two complex components instead of four.

We have no chiral invariance (\ref{ch})\ and the transformation (\ref{chsign}%
) does not apply. For this reason, the signs of the couplings $g_{I}$ cannot
be assumed to be positive. The P and T transformations are modified as
follows: 
\begin{eqnarray}
P_{1} &:&\qquad x_{1}\rightarrow -x_{1},\qquad x_{2,3}\rightarrow
x_{2,3},\qquad \psi _{i}^{I}\rightarrow \gamma _{1}\psi _{i}^{I},\qquad
\sigma \rightarrow -\sigma \ ;  \label{P} \\
T &:&\qquad x_{3}\rightarrow -x_{3},\qquad x_{1,2}\rightarrow x_{1,2},\qquad
\psi _{i}^{I}\rightarrow \gamma _{3}\psi _{i}^{I},\qquad \overline{\psi }%
_{i}^{I}\rightarrow \overline{\psi }_{i}^{I}\gamma _{3},\qquad \sigma
\rightarrow -\sigma .  \nonumber
\end{eqnarray}
The theory is still invariant under the duality (\ref{dual}) and we can
assume that $-1\leq g_{I}\leq 1$.

To avoid the problem of extending to $d$ dimensions a Dirac algebra in which
the trace of an odd number of gamma matrices does not vanish, it is better
to keep $d=3$ and regularize the theory modifying the $\psi $- and $\sigma $%
-quadratic terms with a cut-off as follows: 
\[
\overline{\psi }\partial \!\!\!\slash \left( 1-{\frac{\Box }{\Lambda ^{2}}}%
\right) \psi ,\qquad \mathrm{and}\qquad {\frac{1}{2}}|\sigma (k)|^{2}\sqrt{%
k^{2}}\left( 1+{\frac{k^{2}}{\Lambda ^{2}}}\right) D(g_{I}^{2},r_{I}). 
\]
This regularization framework manifestly preserves the P and T symmetries (%
\ref{P}) (a Pauli-Villars framework does not) and ensures that no $\sigma
^{3}$ term is generated by renormalization. The graphs with an odd number of 
$\sigma $ external legs are finite, but need not vanish.

The wave-function renormalization constants and beta functions are the same
as in (\ref{zetas}) and (\ref{betas}) with $N\rightarrow N/2$. In
particular, the $g_{I}$ beta functions read 
\begin{equation}
\beta _{I}(g)=-\frac{g_{I}}{2}{\frac{\mathrm{d}\ln Z_{I}}{\mathrm{d}\ln \mu }%
}={\frac{8}{3\pi ^{2}N}g}{_{I}(g_{I}^{2}-1)}D(g_{J}^{2},r_{J}).
\label{betascal}
\end{equation}
We see that the fixed points of this model are $\left\{ g_{I}\right\} $= any
string made of $1,0,-1$. These fixed points have the form 
\[
\Psi _{N_{f}}\otimes F_{2}\Sigma _{N_{\sigma },N_{\sigma }^{\prime }}, 
\]
where $N_{f}$ is the number of decoupled fermions (coupling $g_{I}=0$), $%
N_{\sigma }$ is the number of fermions interacting with $\sigma $ with
coupling $g_{I}=1$ and $N_{\sigma }^{\prime }$ is the number of fermions
interacting with $\sigma $ with coupling $g_{I}=-1$. The conformal field
theories $F_{2}\Sigma _{N_{\sigma },N_{\sigma }^{\prime }}$ are defined by
the lagrangian 
\begin{equation}
\mathcal{L}=\sum_{i=1}^{N_{\sigma }}\overline{\psi }_{i}(\partial \!\!\!%
\slash +\lambda \sigma )\psi _{i}+\sum_{i=j}^{N_{\sigma }^{\prime }}%
\overline{\chi }_{j}(\partial \!\!\!\slash -\lambda \sigma )\chi _{j}.
\label{conf2}
\end{equation}
and have symmetry $O(2N_{\sigma })\otimes O(2N_{\sigma }^{\prime })$. The
subscript 2 in $F_{2}\Sigma _{N_{\sigma },N_{\sigma }^{\prime }}$ means that
the spinors are two-component. In general ($r\neq 1$), the conformality of (%
\ref{conf2}) does not seem to be implied by more general principles than the
vanishing of the beta function (\ref{betascal}) at $g=-1$. The symmetries of
the model allow us to construct a local, dimension-3 operator $O$ that does
not vanish using the field equations (for example $\sigma \overline{\psi }%
\psi $) and could appear in the trace anomaly, $\Theta =\beta O$. Instead,
at $r=1$ the two-component spinors $\psi $ and $\chi $ can be grouped into a
four-component spinor and we have $F_{2}\Sigma _{N,N}=F_{4}\Sigma _{N}$,
whose conformality follows from the general arguments of \cite{largeN}.

The $F_2\Sigma _{N_{\sigma },N_{\sigma }^{\prime }}$ theories have also a
duality 
\[
\psi \leftrightarrow \chi ,\qquad \sigma \rightarrow -\sigma \sqrt{r},\qquad
\lambda \rightarrow \lambda /\sqrt{r},\qquad r\rightarrow 1/r,\qquad
N_{\sigma }\leftrightarrow N_{\sigma }^{\prime }, 
\]
where $r=N_{\sigma }^{\prime }/N_{\sigma }$ and $\lambda ^{2}N_{\sigma }=8$.

The order $\mathcal{O}(1/N)$ anomalous dimensions $\gamma _{\psi }^{I}$ and $%
\gamma _{\sigma }$ of the theories $F_2\Sigma _{N,N^{\prime }}$ coincide
with those of the theories $F_2\Sigma _{N+N^{\prime }}$ (and $%
F_4\Sigma_{N/2,N^{\prime}/2}$, $F_4\Sigma_{(N+N^{\prime})/2}$), but the
symmetry groups differ.

It is straightforward to extend the other results of the previous sections
to the flows interpolating between the $F\Sigma _{N,N^{\prime }}$ models.

\section{Scalar models}

\setcounter{equation}{0}

We now consider the scalar models 
\begin{equation}
\mathcal{L}={\frac{1}{2}}\sum_{I=0}^K\sum_{i=1}^{N_I}
\left[(\partial_{\mu}\varphi_i^I)^2+i
\lambda_I\sigma(\varphi_i^I)^2\right]+V_6(\varphi),  \label{intera}
\end{equation}
with symmetry group $\bigotimes_{I=0}^KO(N_I)$, possibly enhanced at the
fixed points. The $\lambda_I$s are not necessarily positive and $%
V_6(\varphi) $ denotes a potential term of the form $\varphi^6$ induced by
renormalization.

\bigskip

\textbf{Scalar $\Sigma _{N}$ conformal field theories.} I define the scalar $%
\Sigma _{N}$ conformal theories by means of the lagrangian 
\begin{equation}
\mathcal{L}={\frac{1}{2}}\sum_{i=1}^{N}\left[ (\partial _{\mu }\varphi
_{i})^{2}+i\lambda _{I}\sigma (\varphi _{i})^{2}\right] ,  \label{confo}
\end{equation}
which I briefly denote with $S\Sigma _{N}$. This theory coincides with the
UV (Wilson-Fischer) fixed point of the $O(N)$ sigma model. In the approach
of \cite{largeN}, the $S\Sigma _{N}$ model can be studied \textit{per se},
with no reference to an RG\ flow generating it as a critical limit. The
construction of the theories $S\Sigma _{N}$ follows the lines of the
construction of the fermionic theories $F\Sigma _{N}$ of \cite{largeN}. A
caveat is that renormalization generates a $\varphi ^{6}$ term. Due to the $%
O(N)$ symmetry, this term can only have the form 
\begin{equation}
\left( \sum_{i=1}^{N}\varphi _{i}^{2}\right) ^{3}  \label{equa}
\end{equation}
and so can be reabsorbed with an (imaginary) translation of the field $%
\sigma $. This is equivalent to say that (\ref{equa}) is ``trivial'' because
proportional to the $\sigma $ field equation. The $\sigma $ field equation
formally sets the composite operator $\varphi ^{2}$ to zero, but does not
trivialize the theory. More details on the use of the $\sigma $ field
equation can be found in the appendix.

The theory is classically conformal and we choose the classically conformal
subtraction scheme, in which the renormalized $\varphi ^{2}$- and $\varphi
^{4}$-terms are set to zero by default.

The $\sigma $-propagator is generated by the one-loop scalar bubble, which
is proportional to $(k^{2})^{d/2-2}$. If unmodified, a $\sigma $-propagator $%
\sim (k^{2})^{2-d/2}$ generates $\Gamma [0]$s at the subleading orders. It
is therefore necessary to add an evanescent, RG-invariant, non-local term to
the lagrangian and convert the $\sigma $-propagator to $\sqrt{k^{2}}$. This
means, in particular, that the field $\sigma $ has dimension $\left(
d+1\right) /2$ and $\lambda _{\mathrm{B}}=\lambda \mu ^{\varepsilon /2}$.
The $O(1/N)$ diagrams are the same as in Fig. 1. The diagram ($d$) does not
vanish identically, but is finite. Finally, the $O(1/N)$ results for $\gamma
_{\varphi }$ and $\gamma _{\sigma }$ are equal to those of the $F\Sigma _{N}$
models (\ref{unit}), with the replacement $N\rightarrow N/4$.

\bigskip

\textbf{The running $S\Sigma_N$ models.} Let us now consider the theories (%
\ref{intera}). We can still use the classically conformal substraction
scheme, and therefore assume that the terms of the form $\varphi^2$ and $%
\varphi^4$ are subtracted away by default. However, nothing forbids the
generation of terms of the form $\varphi^6$ and this time they cannot be
removed with a translation of the field $\sigma$. The reason is that $\sigma$
multiplies 
\[
\sum_{I=0}^K\lambda_I(\varphi^I)^2, 
\]
but the $\varphi^6$ divergences can be proportional to 
\[
\left(\sum_{I=0}^K\alpha_I^m(\varphi^I)^2\right)
\left(\sum_{I=0}^K\alpha_J^n(\varphi^J)^2\right)
\left(\sum_{I=0}^K\alpha_L^p(\varphi^L)^2\right) 
\]
with $m$, $n$ and $p$ generic. It is therefore necessary to introduce the
most general $\varphi^6$-vertices in the lagrangian, multiplied by
independent couplings.

For simplicity, I focus on a model with two sets of scalar fields, $\varphi
_{i}$, $i=1,\ldots N$ and $\chi _{j}$, $j=1,\ldots Nr$, $K=2$. The
lagrangian reads 
\begin{equation}
\mathcal{L}={\frac{1}{2}}\sum_{i=1}^{N}(\partial _{\mu }\varphi _{i})^{2}+{%
\frac{1}{2}}\sum_{j=1}^{Nr}(\partial _{\mu }\chi _{j})^{2}+i{\frac{\lambda }{%
2}}\sigma \left( \sum_{i=1}^{N}\varphi _{i}^{2}+g\sum_{j=1}^{Nr}\chi
_{j}^{2}\right) +{\frac{32\lambda _{60}}{N^{2}}}\left( \sum_{i=1}^{N}\varphi
_{i}^{2}\right) ^{3}+{\frac{32\lambda _{06}}{N^{2}r^{2}}}\left(
\sum_{j=1}^{Nr}\chi _{j}^{2}\right) ^{3}.  \label{lagra6}
\end{equation}
The factors in front of the $V_{6}$-vertices are chosen for notational
convenience.

The duality transformation is 
\begin{equation}
g\rightarrow {\frac{1}{g}},\qquad \varphi\leftrightarrow\chi,\qquad
r\rightarrow {\frac{1}{r}},\qquad \lambda_{60}\leftrightarrow\lambda_{06},
\qquad \lambda\rightarrow {\frac{\lambda}{\sqrt{r}}}, \qquad
\sigma\rightarrow \sigma g\sqrt{r}.  \label{dual6}
\end{equation}

I have dropped vertices of the form $\varphi ^{4}\chi ^{2}$ and $\varphi
^{2}\chi ^{4}$, since they can always be reabsorbed with a $\sigma $%
-translation while keeping duality manifest. The remaining $V_{6}$-vertices
are redundant (one suffices), but cannot be reduced to a single vertex in
the full space of parameters using the $\sigma $-field equation. We have to
divide the parameter space into two charts, one contaning $g=0$ and the
other containing $g=\infty $. In the first chart, the $\varphi ^{6}$ vertex
is trivial and can be reabsorbed into a $\sigma $-translation. However, the $%
\chi ^{6}$ vertex cannot be reabsorbed into a $\sigma $ translation at the
point $g=0$. Therefore, in the first chart we remove the $\varphi ^{6}$
vertex setting $\lambda _{60}=0$ and take $\lambda _{06}$ as the unique $%
V_{6}$-coupling. In the second chart, which contains $g=\infty $, we have a
symmetric situation: the $\chi ^{6}$ vertex can be reabsorbed away, but the $%
\varphi ^{6}$ vertex cannot, so we set $\lambda _{06}=0$ and the true
coupling is $\lambda _{60}$.

In each chart we have an appropriate specialization of the duality
transformation (\ref{dual6}). For example, when we apply duality in the
first chart, we transform the potential $\lambda_{60}\varphi^6/N^2$ into $%
\widetilde{\lambda}_{60}\chi^6/(N^2 r^2)$. Then, we can use the $\sigma$%
-field equation $\chi^2=-\varphi^2/g$ and rewrite the potential as $-%
\widetilde{\lambda}_{60}\varphi^6/(r^2g^3N^2)$. To recover the initial
potential we define $\widetilde{\lambda}_{60}=-r^2g^3\lambda_{60}$, so that
duality in the first chart reads 
\begin{equation}
g\rightarrow {\frac{1}{g}},\qquad \varphi\leftrightarrow\chi,\qquad
r\rightarrow {\frac{1}{r}},\qquad
\lambda_{60}\rightarrow-r^2g^3\lambda_{60}, \qquad \lambda\rightarrow {\frac{%
\lambda}{\sqrt{r}}}, \qquad \sigma\rightarrow \sigma g\sqrt{r}.
\label{first}
\end{equation}
A symmetric argument allows us to specialize the duality transformation to
the second chart. Invariance under duality is a powerful tool to check the
calculations.

\bigskip

\textbf{The leading and subleading order in the $\sigma $-sector.} The
effective $\sigma $-propagator is obtained resumming the scalar bubbles and
reads 
\[
\langle \sigma (k)\,\sigma (-k)\rangle =\sqrt{k^{2}}D(\alpha _{J},r_{J}). 
\]

To the order $\mathcal{O}(1/N)$, the $\varphi ^{6}$ and $\chi ^{6}$ vertices
have no effect on the renormalization of the scalar kinetic terms and the $%
\sigma $-scalar-scalar vertices. This can be proved observing that the
relevant graphs are in one-to-one correspondence with the graphs of Fig. 4
contaning four scalar external legs exiting from the same six-leg vertex. I\
call these graphs $\mathcal{G}_{4,2}$ for clarity.

$i$) The graphs contributing to the renormalization of the scalar propagator
are obtained from $\mathcal{G}_{4,2}$ suppressing the four external scalar
legs exiting from the same six-leg vertex. We obtain the graph (b) of Fig.1
plus tadpoles, finite graphs and a graph with a scalar bubble (from the
third of Fig. 4). This graph should not be counted, since the scalar bubbles
are resummed into the $\sigma $-propagator.

$ii$) The graphs contributing to the renormalization of the $\sigma $%
-scalar-scalar vertices are obtained from $\mathcal{G}_{4,2}$ replacing the
four external scalar legs exiting from the same six-leg vertex with one
external $\sigma $ leg. We obtain the graphs (c) and (d) of Fig. 1 plus
tadpoles and finite graphs.

In summary, the order $\mathcal{O}(1/N)$ anomalous dimensions and beta
functions are exactly the same as in (\ref{betas}), with $N\rightarrow N/4$.
In particular, we have 
\[
\beta _{g}={\frac{32}{3N\pi ^{2}}}{\frac{g(g^{2}-1)}{1+rg^{2}}}. 
\]
The zeroes are at $g=-\infty ,-1,0,1,\infty $.

\bigskip

\textbf{The $\varphi ^{6}$-terms.} Now we study the induction of a $V_{6}$%
-coupling by renormalization. The order $\mathcal{O}(1/N)$ beta function of
the $\varphi ^{6}$-coupling reads in the first chart 
\begin{eqnarray}
\beta _{60} &=&{\frac{32}{3N\pi ^{2}(1+rg^{2})^{3}}}\left[
-(1-g)^{3}+3\lambda _{60}\left( 1+3g^{2}+2rg^{2}+6rg^{3}+4r^{2}g^{4}\right)
\right.  \nonumber \\
&&\left. \qquad \qquad \qquad \qquad \qquad \qquad -27\lambda
_{60}^{2}rg^{3}(1-r^{2}g^{3})-27\lambda _{60}^{3}r^{3}g^{6}\right] .
\label{beta60}
\end{eqnarray}
Graphs and details on the calculation are given in the appendix. This
formula transforms correctly under the duality (\ref{first}), i.e. 
\begin{equation}
\beta _{60}\rightarrow -r^{2}g^{3}\beta _{60}-3\lambda _{60}r^{2}g^{2}\beta
_{g}.  \label{betadual}
\end{equation}
In the second chart we have 
\begin{eqnarray*}
\beta _{06} &=&{\frac{32}{3Nr\pi ^{2}(1+rg^{2})^{3}}}\left[
r^{3}g^{3}(1-g)^{3}+3\lambda _{06}rg^{2}\left(
4+6rg+2rg^{2}+3r^{2}g^{2}+r^{2}g^{4}\right) \right. \\
&&\left. \qquad \qquad \qquad \qquad \qquad \qquad +27\lambda
_{06}^{2}(1-r^{2}g^{3})-27\lambda _{06}^{3}\right] .
\end{eqnarray*}

\medskip

\textbf{Fixed points.} At $g=\infty $ and $g=0$ we work in the first and
second chart, respectively, and find the beta function of ref.s \cite
{pisarskietal}, namely 
\begin{equation}
\beta _{60}={\frac{288}{N\pi ^{2}}}\lambda _{60}^{2}(1-\lambda _{60}),\qquad
\beta _{06}={\frac{288}{Nr\pi ^{2}}}\lambda _{06}^{2}(1-\lambda _{06}).
\label{beta0}
\end{equation}
The solutions of $\beta _{60}=0$ are $\lambda _{60}^{*}=0$ and $\lambda
_{60}^{*}=1$. At $\lambda _{60}^{*}=0$ the fixed point is the direct product
of an $S\Sigma _{Nr}$ model and $N$ free scalars. At $\lambda _{60}=1$ the
fixed point is the direct product of $S\Sigma _{Nr}$ and the interacting $%
\varphi _{3}^{6}$ fixed point. In both cases, the flavor symmetry group is $%
O(N)\otimes O(Nr)$. The negative values of $\lambda _{60}$ and $\lambda
_{60} $ are unphysical at $g=\infty $ and $g=0$, becuse the potential is
unbounded from below.

For $g=1$ we have 
\[
\beta _{60}={\frac{32\lambda _{60}}{N\pi ^{2}(1+r)^{3}}}\left[
4(1+r)^{2}-9\lambda _{60}r(1-r^{2})-9\lambda _{60}^{2}r^{3}\right] . 
\]
The solutions of $\beta _{60}=0$ are $\lambda _{60}^{*}=0$ and the two
interacting fixed points 
\[
\lambda _{60}^{*}={\frac{1+r}{6r^{2}}}\left( 3(r-1)\pm \sqrt{9-2r+9r^{2}}%
\right) , 
\]
which are interchanged by duality. At $\lambda _{60}=0$ the fixed point is
an $S\Sigma _{N(1+r)}$ model.

For $g=-1$ we have 
\[
\beta _{60}={\frac{32}{3N\pi ^{2}(1+r)^{3}}}\left[ -8+12\lambda
_{60}(1-r+r^{2})+27\lambda _{60}^{2}r(1+r^{2})-27\lambda
_{60}^{3}r^{3}\right] . 
\]
The equation $\beta _{60}=0$ has three real solutions, two positive and one
negative. For $r\rightarrow \infty $ the largest positive solution tends to $%
1$ from above, the second positive solution $\bar{\lambda}_{60}$ tends to $%
0^{+}$ and the negative solution tends to $0^{-}$.

The fixed point $(g,\lambda _{60})=(0,1)$ is UV unstable, while the fixed
points $(1,0)$ and $(-1,\bar{\lambda}_{60})$ are IR stable. The other fixed
points are partially stable and partially unstable.

\begin{figure}[tbp]
\centerline{\epsfig{figure=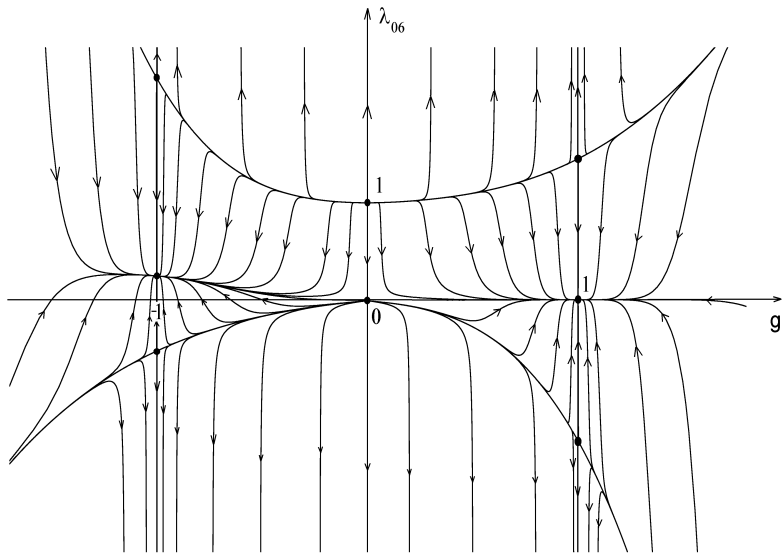,height=9cm,width=12cm}}
\caption{The fixed points and RG flows of the scalar models}
\end{figure}

The RG flows are plotted in Fig. 3. Since at $g=0$ the negative values of $%
\lambda _{60}$ are unphysical, we can argue that the unphysical region is
the region below the line connecting the three fixed points with $\lambda
_{60}^{*}\leq 0$.

A particularly simple case is the limit $r\rightarrow \infty $, where $\beta
_{60}$ is independent of $g$ and coincides with (\ref{beta0}). Here the RG
flows of $g$ and $\lambda _{60}$ separate.

\section{Conclusions}

In the realm of high-energy physics, classically conformal theories play a
special role. Taking inspiration from massless QCD, where hadrons and
glueballs acquire masses dynamically, it is tempting to think that the
ultimate theory of the universe contains no dimensionful parameter of
classical origin, the classically conformal subtraction scheme should be
universally used and the dynamical generation of a scale $\mu $ by
dimensional transmutation originates all scales and dimensionful parameters
of nature, including the Newton constant.

Another context in which the specialness of classically conformal theories
is singled out, at least in even dimensions, is the irreversibility of the
RG flow. It would be interesting to know if the RG flow is irreversible also
in odd dimensions. The flows of this paper are used for this investigation
in ref. \cite{ineq}.

It is useful to have a large class of classically conformal models to test
ideas around the issues just mentioned and the strongly-coupled limit of
quantum field theory. I believe that the models studied in this paper can
help us answering various unsolved problems and have a number of potential
applications, maybe also in condensed-matter physics.

I have explored two classes of theories, describing sets of fermions and
scalar fields interacting by means of a $\sigma $-field, with a variety of
fixed points, known and new. Typically, both end points of the flows are
interacting conformal field theories. The remarkable properties of these
theories are that $i$)\ at each order of the $1/N$ expansion the RG flows
can be integrated for arbitrary values of the running couplings; $ii$)
``exact'' beta functions, anomalous dimensions and correlation functions
interpolating between the UV and IR conformal limits can be written; $iii$)
the models have manifest $g\rightarrow 1/g$ dualities at each order of the $%
1/N$ expansion. Duality puts severe constraints on the formulas and is a
powerful tool to check the calculations. In special limits ($r\rightarrow 0$
in the fermionic models, $r\rightarrow \infty $ in the scalar models) the
integration of the flow equations further simplifies, yet the flow remains
non-trivial.

\vskip .3truecm \textbf{Appendix: diagrammatics and duality in the scalar
models}

\vskip .1truecm

The graphs contributing to the renormalization of $\varphi ^{6}$ to the
order $\mathcal{O}(1/N)$ are shown in Fig. 4. Some diagrams have obvious
counterparts with two, four and six external $\chi $-legs. The divergent
terms proportional to $\varphi ^{2}\chi ^{4}$, $\varphi ^{4}\chi ^{2}$ and $%
\chi ^{6}$ are converted to $\varphi ^{6}$ using the $\sigma $-field
equation $\varphi ^{2}+g\chi ^{2}=0$. (For subtleties in the use of this
field equation see below.) Several diagrams are finite, but have been drawn
for completeness.

\begin{figure}[tp]
\centerline{\epsfig{figure=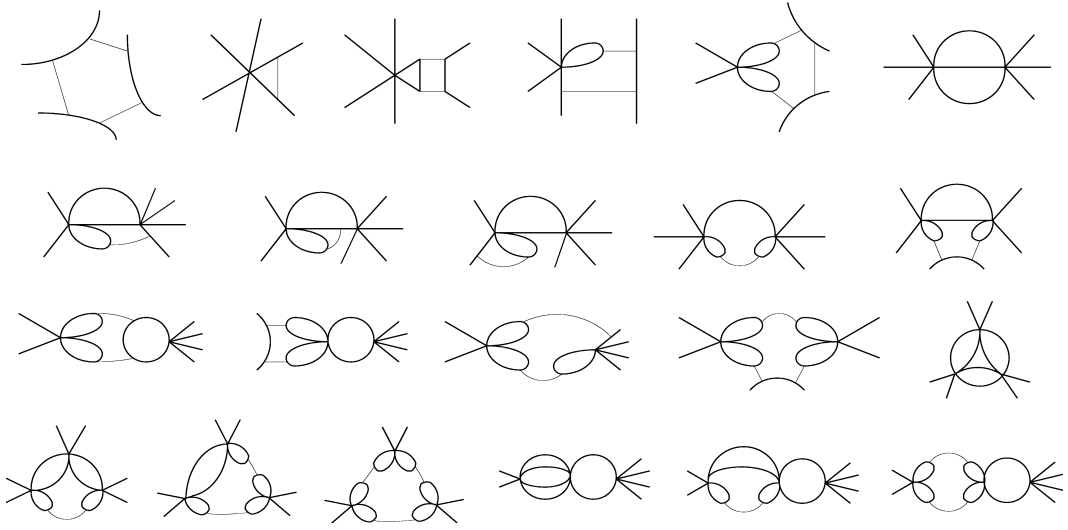,height=8cm,width=15cm}}
\caption{Renormalization of $\varphi^6$ in the scalar models}
\end{figure}

The beta function (\ref{beta60}) is invariant under the duality symmetry (%
\ref{first}) and (\ref{betadual}). However, the renormalization constant $%
Z_{60}$ is not. The complete expression of $\lambda _{60}Z_{60}$ is very
long and contains terms with the first three powers of $\lambda _{60}$ plus
a term independent of $\lambda _{60}$. Here I focus on the terms
proportional to $\lambda _{60}^{3}$, for simplicity, which are 
\begin{equation}
-{\frac{288\lambda _{60}^{3}}{N\pi ^{2}}}\left( {\frac{\mu ^{-4\varepsilon }%
}{4\varepsilon }}-{\frac{3\mu ^{-5\varepsilon }}{5\varepsilon (1+rg^{2})}}+{%
\frac{\mu ^{-6\varepsilon }}{2\varepsilon (1+rg^{2})^{2}}}-{\frac{\mu
^{-7\varepsilon }}{7\varepsilon (1+rg^{2})^{3}}}\right) .  \label{mis}
\end{equation}
Expanding $\mu ^{-m\varepsilon }/(m\varepsilon )$ as $1/(m\varepsilon )-\ln
\mu $ we see that the terms proportional to $\ln \mu $, which contribute to
the beta function, transform correctly under duality, while the pole terms,
which have no physical significance, do not transform correctly.

These facts are explained as follows. The symmetry of the theory under the
duality transformation (\ref{first}) in the first chart understands the use
the $\sigma $ field equation. Naively, the $\sigma $ field equation is $%
\varphi ^{2}+g\chi ^{2}=0$, but the modified regularization technique of 
\cite{largeN} demands to add the non local term 
\[
\frac{1+rg^{2}}{2}\int {\frac{\mathrm{d}^{d}k}{(2\pi )^{d}}}|\sigma (k)|^{2}{%
\frac{1}{\sqrt{k^{2}}}}\left( 1-{\frac{\lambda _{\mathrm{B}}^{2}N}{16}}%
\left( k^{2}\right) ^{-\varepsilon /2}\right) 
\]
to the action $S$. The complete $\sigma $ field equation reads, in momentum
space, 
\[
\varphi ^{2}(k)+g\chi ^{2}(k)={\frac{2i(1+rg^{2})}{\lambda }}{\frac{1}{\sqrt{%
k^{2}}}}\left( 1-{\frac{\lambda _{\mathrm{B}}^{2}N}{16}}\left( k^{2}\right)
^{-\varepsilon /2}\right) \sigma (k). 
\]
The evanescent right-hand side of this formula provides extra vertices which
do contribute to the pole terms. However, these corrections are always
proportional to 
\begin{equation}
{\frac{\mu ^{-m_{1}\varepsilon }}{m_{1}\varepsilon }}-{\frac{\mu
^{-m_{2}\varepsilon }}{m_{2}\varepsilon }}  \label{form}
\end{equation}
for some $m_{1}$ and $m_{2}$ and so do not affect the $\ln \mu $ terms. This
fact can be easily verified in a number of examples. Moreover, the terms
proportional to the $\sigma $ field equation $\delta S/\delta \sigma $
cannot be neglected. They give other contributions of the form (\ref{form}),
which can be easily computed functionally integrating by parts. Putting
these ingredients together, we discover that the mismatch between $Z_{60}$
and its dual is due to the use of the $\sigma $ field equation and the
difference\ between the naive and regularized $\sigma $ field equations. It
is not a lengthy work to make a quantitative check for the terms (\ref{mis}%
). The important point is that the mismatch has no physical significance. We
can neglect these nuisances if we limit ourselves to check duality within
the physical quantities.

\vskip .5truecm \textbf{Acknowledgements}

\vskip .2truecm

I am grateful to the Aspen Center for Physics for warm hospitality during
the realization of this work and P. Calabrese, P. Parruccini and E. Vicari
for discussions on three-dimensional models and critical phenomena.

\end{document}